\documentclass[a4paper,preprintnumbers,showpacs,twocolumn,superscriptaddress,nofootinbib,amsmath,amssymb]{revtex4-1}
\usepackage{amsmath,amssymb,graphicx,amsfonts}
\usepackage{mathrsfs}
\usepackage{comment}
\usepackage{xcolor}
\usepackage[shortlabels]{enumitem}
\usepackage[unicode, colorlinks=true, linkcolor=magenta, citecolor=cyan, filecolor=cyan, urlcolor=blue, linktocpage, breaklinks]{hyperref}
\usepackage{url}
\usepackage{graphicx}
\usepackage{dcolumn}
\usepackage{bm}
\usepackage{wasysym}
\usepackage{color}
\usepackage{multirow}
\usepackage{booktabs}

\begin{document}
\title{Optimal colliding energy for the synthesis of a superheavy element $Z$=119}

\author{Avazbek Nasirov}
\email{nasirov@jinr.ru}
\affiliation{Joint Institute for Nuclear Research, Dubna 141980, Russia}
\affiliation{Institute of Nuclear Physics, Tashkent 100214, Uzbekistan}
\author{Bakhodir Kayumov}
\email{b.kayumov@newuu.uz} 
\affiliation{School of Humanities and Natural Sciences, New Uzbekistan University, Tashkent 100007, Uzbekistan}
\affiliation{Institute of Nuclear Physics, Tashkent 100214, Uzbekistan}

\begin{abstract}
The evaporation residue (ER) cross sections of 3n and 4n channels related to the synthesis of superheavy element (SHE) with the charge number $Z=119$ in the $^{51}$V+$^{248}$Cm reaction have been calculated by the dinuclear system (DNS) model as a sum of the partial cross sections of the corresponding channels. The angular momentum distribution of the compound nucleus (CN) is estimated by the dynamical trajectory calculations of the capture probability which is considered as the DNS formation probability. The fusion probability decreases by the increase of the DNS angular momentum due to its influence on the intrinsic fusion barrier $B_{\rm fus}^*$. The range $\alpha_2=60^{\circ}-70^{\circ}$ of the orientation angle of the axial symmetry axis of the deformed target nucleus $^{248}$Cm is favorable for the formation of the CN. The fusion probability decreases at around $\alpha_2=90^{\circ}$ since the number of the partial waves contributing to the capture decreases. Therefore, it is important to calculate the capture cross section dynamically. The 4n channel cross section of the SHE synthesis is larger than the 3n channel cross section maximum value of the ER cross section and it is 12.3 fb at $E_{\rm c.m.}$=232 MeV.
\end{abstract}

\maketitle
\section{Introduction}

The synthesis of superheavy elements (SHE) represents a remarkable achievement in the field of nuclear physics \cite{Hofman2000, Giuliani2019}. These exotic elements, characterized by staggering atomic masses and short lifetimes, have intrigued scientists for decades. The heaviest superheavy element which was synthesized by using $^{48}$Ca-induced complete fusion reactions have been successfully used to synthesize SHE with charge numbers $Z=$112–118 in the neutron evaporation channels \cite{Stavsetra2009, Oganessian2010, Dullmann2010, Oganessian2013, Oganessian2013_2, Khuyagbaatar2014, Utyonkov2015}. However, reactions $^{50}$Ti$+^{249}$Bk and $^{249}$Cf \cite{Khuyagbaatar2020},  $^{54}$Cr$+^{248}$Cm \cite{Hofmann2016} and $^{58}$Fe$+^{244}$Pu \cite{Oganessian2009} did not provide success for the synthesis of SHEs with $Z=119$ and $120$.

A lot of theoretical investigations devoted to describe the measured data of the evaporation residue cross sections and to making predictions for their values in the different reactions leading to the formation of the SHE with $Z=119$ and $120$ have been done \cite{Zhu2014, Li2018, Bao2015, Liu2013, Ghahramany2016, Adamian2018, Adamian2020, Li2022, Lv2021, Wilczynska2019}. Their results can be useful to make a choice of a pair of colliding nuclei and to fix a range of the beam energies leading to the observable cross sections in the synthesis of the wanted SHE. The authors in Ref. \cite{Li2018} made a conclusion that the most promising reactions of the synthesis of superheavy elements $Z=$119 and $Z=$120 are $^{50}$Ti + $^{248}$Bk and $^{50}$Ti + $^{249,251}$Cf, respectively. For example, for the synthesis of the superheavy element $Z=$119, values of the ER cross section in the $^{50}$Ti+$^{249}$Bk and $^{51}$V+$^{248}$Cm reactions are predicted in the range 1$-$100 fb \cite{Zhu2014, Ghahramany2016, Adamian2018, Adamian2020, Li2022, Lv2021, Wilczynska2019} (the detail information is presented on page 8 of this work).

The very small cross sections obtained in the synthesis of SHE by different reactions require finding a favorable reaction (projectile and target pair) among them and the optimal beam energy range which is very narrow (no more than $10$ MeV). The ER formation process is the last stage of the complicated reaction mechanism in heavy-ion collisions near the Coulomb barrier energies. Processes, that can held during the formation of the ER, and can be in competition with the ER, are described below (also can be seen in Fig.1 \cite{Kayumov2022}):
\begin{enumerate}
\item The initial phase includes a competition between deep-inelastic collision where incomplete momentum transfer occurs and the capture process with the full momentum transfer. Depending on the values of initial energy and angular momentum, one of these two processes can held. As the energy of the incident nucleus decreases, the projectile can stuck in a potential well, and this is the capture of the projectile by the target nucleus. However, the relative kinetic energy of the system can be large enough to overcome the Coulomb barrier, even after the dissipation of sufficient part, and this process is called deep-inelastic collision (see Fig. 2 in Ref.~\cite{Nasirov2005}). As mentioned above, competition between these processes strongly depends on the incident energy of the system and the angular momentum of relative motion. Capture is the first step towards the formation of a superheavy element and it leads to the formation of a molecular-like dinuclear system with a larger lifetime. During this time it can develop, changing its charge and mass asymmetry due to multinucleon transfer and changing its form.

\item Competition between the quasifission process and the formation of the excited mononucleus with the large angular momentum, is the next stage in the evolution of the DNS. After capturing the projectile nucleus by target, DNS may break up into two fragments, and the products of this mechanism are called quasifission products. Therefore, complete dissipation of kinetic energy (capture stage) is the main characteristic of the quasifission process. It strongly depends on the excitation energy $E^*_{\rm DNS}$ and angular momentum $L=\ell\hbar$ of the dinuclear system, nuclear shapes (deformations), and the orientation angles of the nuclei relative to the beam direction. In addition, the quasifission process is the main hindrance to the evolution of the DNS into the CN.

\item In the next phase of the process, a rotating and excited mononucleus, which is formed by the transfer of all nucleons from lighter to heavier nuclei, should reach the equilibrium state to form the  CN. During this evolution, it can decay into two fragments, and the products of this splitting are named fast fission. As was discussed in Ref. \cite{Sierk1986}, the angular momentum of the excited mononucleus has an inverse proportionality to the fission barrier. The range of the angular momentum of the system is established in the capture stage. This means mononucleus with the larger angular momentum has a lower fission barrier or no fission barrier entirely. In such kind of scenarios, the excited mononucleus will go into the fast fission instead of going to the equilibrium state.
\end{enumerate}

Finally, the formed CN can break up which gives fusion-fission products, or can survive against fission by cooling (emission of neutrons, protons, alpha particles, and gamma quanta). This competition strongly depends on the excitation energy of the system and the fission barrier, and it can be calculated by the survival probability $W_{\rm sur}$. According to all these stages, ER cross section $\sigma_{\rm ER}$ can be calculated as summation over all partial waves $\ell$ as described in references \cite{ADAMIAN200024, Nasirov2011}:
\begin{eqnarray}\label{ERcs}
\sigma_{\rm ER}(E_{\rm c.m.})&=&\sum_{\ell} \sigma_{\rm cap}(E_{\rm c.m.},\ell) P_{\rm CN}(E_{\rm c.m.},\ell) \nonumber  \\
&\times& W_{\rm sur}(E_{\rm c.m.},\ell)
\end{eqnarray}
The short description of the model used in the calculation of the cross sections of the capture, complete fusion, and ER formation in competition with the fission is presented in Section \ref{model}. The results of the calculation of the partial fusion, quasifission and ER cross sections are discussed in Section \ref{discuss}.

\section{Model}
\label{model}
\subsection{Capture cross section}
Two conditions must be satisfied for the capture in the collision dynamics.\\ 
1) The initial energy $E_{\rm c.m.}$ of a projectile in the center-of-mass system should be larger than the minimum of the potential well of the nucleus-nucleus interaction (Coulomb barrier + rotational energy of the entrance channel) to overcome or by tunneling through the barrier along relative distance in the entrance channel to form a DNS.\\
2) At the same time the value of the relative kinetic energy above the entrance channel barrier should be in correspondence with the size of the potential well: in case of the collision of the massive nuclei the size of the potential is small and, if the initial collision energy is very large relative to the entrance channel barrier, the dissipation of the kinetic energy maybe not enough to make its value lower than a barrier of potential well,  i.e. to cause trapping into potential well. As a result, the capture does not occur and the deep-inelastic collision takes place. These events are separated by the analysis of the collision dynamics. The range of the values of the orbital angular momentum leading to capture is determined by solving the equations of motion for the relative distance $R$ and orbital angular momentum $\ell$ \cite{Fazio2003,Nasirov2005}. Trajectory calculations show that the values of $\ell$ leading to capture can form a "window". The boundary values $\ell_m$ and $\ell_d$ of the $\ell$-"window" leading to capture depend on the beam energy, the size of the potential well of the nucleus-nucleus potential, and the friction coefficients of radial motion and angular momentum. The values of $\ell_m$ and $\ell_d$ are found from calculation of the collision trajectory of nuclei by the solution of the equation of motion for the radial  distance $R$ between their centres-of-mass and orbital angular momentum  \cite{Fazio2003,Nasirov2005}. The capture probability $\mathcal{P}^{\ell}_{cap}(E_{\rm c.m.},\ell;\{\alpha_i,\beta_i\})$ for the collision energy $E_{\rm c.m.}$ depends on the orientation angles $\alpha_i$ of the axial symmetry axis of the colliding deformed nuclei $i=1$ and 2; $\beta_i$ are their deformation parameters. The capture probability is determined from the trajectory calculations by the use of following conditions:
\begin{widetext}
\begin{eqnarray}
\mathcal{P}^{(\ell)}_{cap}(E_{\rm c.m.},\ell,\{\alpha_i,\beta_i\}) = \left\{
\begin{array}{lll}
1,\,\, \mbox{if}\,\, \ell_m < \ell < \ell_d \,\, \mbox{and} \,\, E_{\rm c.m.} > V_{B}, \\
0,\,\, \mbox{if} \,\,\ell < \ell_m \,\, \mbox{or}\,\,\ell>\ell_d \,\, \mbox{and} \,\, E_{\rm c.m.}>V_{B}\, \\
\mathcal{P}^{(\ell)}_{tun}(E_{\rm c.m.},\ell,\{\alpha_i,\beta_i\}),\,\, \mbox{for all}\,\, \ell \mbox{ if } V_{\rm min} < E_{\rm c.m.}\leq V_{B},\, \\
\end{array} \right.
\label{CapClass}
\end{eqnarray}
\end{widetext}
where $V_B$ and $V_{\rm min}$ are the barrier and minimum values of the potential well of the nucleus-nucleus interaction in the entrance channel. 

Theoretical values of the capture cross sections are calculated with the quantities characterizing the entrance channel by the formula \cite{Kayumov2022}:
\begin{eqnarray}\label{capture}
\sigma_{\rm cap}(E_{\rm c.m.},\ell;\{\alpha_i,\beta_i\})&=&\frac{\lambda^2}{4\pi}(2\ell+1)\nonumber\\
&\times&\mathcal{P}^{\ell}_{\rm cap}(E_{\rm c.m.},\ell;\{\alpha_i,\beta_i\}),
\end{eqnarray}
where $\lambda$ is the de Broglie wavelength of the entrance channel.  
$\mathcal{P}^{(\ell)}_{tun}$ is the probability of the barrier penetrability which is calculated by the improved WKB formula by Kemble  \cite{Kemble1935}:
\begin{eqnarray}\label{penetration}
\mathcal{P}^{(\ell)}_{tun}(E_{\rm c.m.},\{\alpha_i,\beta_i\})=\frac{1}
{1+\exp\left[2K(E_{\rm c.m.},\ell,\{\alpha_i,\beta_i\})\right]},\nonumber\\
\end{eqnarray}
where

\begin{eqnarray}
&&K(E_{\rm c.m.},\ell,\{\alpha_i,\beta_i\})\nonumber\\
&&=\int\limits_{R_{in}}^{R_{out}} dR\times\sqrt{\frac{2\mu}{\hbar^2}(V(R,\ell,\{\alpha_i,\beta_i\})-E_{\rm c.m.})}.
\end{eqnarray}
$R_{in}$ and $R_{out}$ are inner and outer turning points which were estimated by $V(R)=E_{\rm c.m.}$.

\subsection{Fusion cross section}
The partial fusion cross section is determined by the product of capture cross section $\sigma_{\rm cap}(E_{\rm c.m.},\ell,\{\alpha_i,\beta_i\})$ and the fusion probability $P_{CN}(E_{\rm c.m.},\ell,\{\alpha_i,\beta_i\})$ of DNS for the various excitation energies \cite{Kim2015, Fazio2005}:
\begin{eqnarray}\label{fusion}
\sigma_{\rm fus}(E_{\rm c.m.},\ell;\{\alpha_i,\beta_i\})&=&\sigma_{\rm cap}(E_{\rm c.m.},\ell;\{\alpha_i,\beta_i\})\nonumber\\
&\times& P_{\rm CN}(E_{\rm c.m.},\ell;\{\alpha_i,\beta_i\}).
\end{eqnarray}

The fusion probability $P_{CN}(E_{\rm c.m.},\ell,\{\alpha_i,\beta_i\})$ takes into account the change in mass and charge distributions of $D_Z$ in DNS fragments after capture \cite{Kayumov2022}. In general, it is calculated as the sum of the competing channel of quasifission and complete fusion at different charge asymmetries from the symmetric configuration $Z_{sym}$ of the DNS to the configuration corresponding to the maximum value of the driving potential $Z_{max}$ and can be represented in this form:
\begin{eqnarray}
P_{\rm CN}(E_{\rm c.m.},\ell,\{\alpha_i,\beta_i\})
&=&\sum\limits^{Z_{max}}_{Z_{sym}}D_Z(E^*_Z,\ell,\{\alpha_i,\beta_i\})\nonumber\\
&\times& 
P^{(Z)}_{\rm CN}(E^*_Z,\ell,\{\alpha_i,\beta_i\}).
\label{PcnDZ}
\end{eqnarray}
The values of the $D_Z(E^*_Z,\ell,\{\alpha_i,\beta_i\})$ are calculated by the solution of the transport master equation with the nucleon transition coefficients depending on the occupation numbers and energies of the single-particle states of nucleons of the DNS nuclei \cite{Kayumov2022}. The fusion probability $P^{(Z)}_{CN}(E^*_Z,\ell,\{\alpha_i,\beta_i\})$ for the DNS fragments with the charge configuration $Z$ rotating with the orbital angular momentum $\ell$ is calculated as the branching ratio of the level densities of the quasifission barrier $B^Z_{qf}$ at a given mass asymmetry, over the intrinsic barrier $B^{*(Z)}_{fus}$ and symmetry barrier $B^{(Z)}_{\rm sym}$ on mass asymmetry axis  \cite{Nasirov2019}:
\begin{equation}
 \label{Pcnro} P^{(Z)}_{CN}(E^*_{Z})=\frac{\rho_{\rm fus}(E^*_{Z})
}{\rho_{\rm fus}(E^*_{Z}) +
\rho_{\rm qfiss}(E^*_{Z})+\rho_{\rm sym}(E^*_{Z})}.
 \end{equation}
The use of the level density function of the Fermi system leads to the formula for the fusion probability at the DNS excitation energy $E^*_Z$ and angular momentum $L$ from its charge asymmetry 
$Z$:
\begin{equation}
 \label{Pcn} P^{(Z)}_{CN}(E^*_Z)=\frac{e^{-B_{\rm fus}^{*(Z)}/T_Z}}{e^{-B_{\rm fus}^{*(Z)}/T_Z} +
e^{-B_{\rm qfiss}^{*(Z)}/T_Z}+e^{-B_{\rm sym}^{*(Z)}/T_Z}}.
 \end{equation}
Here the values of the level density on the barriers  $B^{(Z)*}_{\rm fus}(\alpha_i,\beta_i)$, $B^{*(Z)}_{\rm sym}(\alpha_i,\beta_i)$ and $B^{(Z)}_{qf}(\alpha_i,\beta_i)$ have been used. To simplify the presentation of Eqs. (\ref{Pcnro})  and (\ref{Pcn}) the arguments $(\alpha_i,\beta_i)$ of the functions $E^*_Z(\alpha_i,\beta_i)$, $T_Z(\alpha_i,\beta_i)$, $B^{*(Z)}_{\rm fus}(\alpha_i,\beta_i)$, $B^{*(Z)}_{\rm sym}(\alpha_i,\beta_i)$ and $B^{(Z)}_{qf}(\alpha_i,\beta_i)$ are not indicated on their right sides of Eqs. (\ref{Pcnro})  and (\ref{Pcn}).  The excitation energy $E^*_Z(E_{\rm c.m},\ell)$ of the DNS with the charge $Z$ and mass $A$ numbers of the light fragment is determined by the difference between collision energy $E_{\rm c.m}$ and peculiarities of the driving potential $U_{dr}$ calculated for the given value of $\ell$:
\begin{eqnarray}
E^*_Z(E_{\rm c.m},\ell,\alpha_i,\beta_i)&=&E_{\rm c.m}-V_{\rm min}(Z_P,A_P,R_m,\alpha_i,\beta_i)\nonumber\\
&+&\Delta U(Z,A,\ell,\alpha_i,\beta_i),
\label{ExiZ}
\end{eqnarray}
where
\begin{eqnarray}
\Delta U(Z,A,\ell,\alpha_i,\beta_i)&=&U(Z_P,A_P,\ell,\alpha_i,\beta_i)\nonumber\\
&-&U(Z,A,\ell,\alpha_i,\beta_i)
\end{eqnarray}
is a change of the driving potential of the DNS during its evolution from the initial value ($Z=Z_P$ and $A=A_P$):
\begin{eqnarray}
U(Z_P,A_P,\ell,\alpha_i,\beta_i)&=&B_P(Z_P,A_P)+B_T(Z_T,A_T)\nonumber\\
-B_{\rm CN}&+&V_{\rm min}(Z_P,A_P,\ell,\alpha_i,\beta_i)
\end{eqnarray}
to the final configuration with the  charge and mass numbers $Z$ and $A$, respectively:
\begin{eqnarray}
U(Z,A,\ell,\alpha_i,\beta_i)&=&B_1(Z,A)+B_2(Z_{\rm tot}-Z,A_{\rm tot}-A)\nonumber\\
&-&B_{\rm CN}+V_{\rm min}(Z,A,\ell,\alpha_i,\beta_i),
\end{eqnarray}
where $Z_{\rm tot}=Z_P+Z_T$ and $A_{\rm tot}=A_P+A_T$.
When the final configuration of the DNS is the CN, Eq. (\ref{ExiZ}) gives the CN excitation energy:
\begin{equation}
E^*_{\rm CN}(E_{\rm c.m},\ell)=E_{\rm c.m}+Q_{gg}-V_{\rm CN}(\ell),
\label{Ecn}
\end{equation}
where $Q_{gg}=B_1+B_2-B_{\rm CN}$ is the energy balance of the reaction, $B_1,~B_2$ and $B_{CN}$ are the binding energies of the interacting nuclei and  CN, which are taken from the table in Ref. \cite{AUDI2003, MolNix1995}; $V_{\rm CN}(\ell)$ is the CN rotational energy.

 The parameters of the quadrupole and octupole deformations of the ground states are taken from Ref. \cite{MolNix1995} of the reacting nuclei in this work while the ones for the first excited $2^+$ and $3^-$ states are obtained from Refs.  \cite{Raman2001} and  \cite{Spear1989}, respectively. The target nucleus $^{248}$Cm has a deformed shape in the ground state (see Table \ref{defpara}) and possibilities of the collision with the different orientation angles $\alpha_2$ relative to the beam direction are taken into account by averaging the contributions of collisions with different values of $\alpha_2$:
\begin{eqnarray}
\langle \sigma_{\rm fus}(E_{\rm c.m},{\beta^{(1)}_i},\ell) \rangle_{\alpha_2} &=& \int_0^{\pi /2} \sin\alpha_2 d\alpha_2\nonumber\\
&\times&\sigma_i (E_{\rm c.m},\ell;{\beta^{(1)}_i},\alpha_2) .
\label{orien}
\end{eqnarray}
The projectile nucleus $^{51}$V is spherical in its ground state (see Table \ref{defpara}) but the first excited quadrupole state $\beta_{2+}=0.2$ is considered as the zero-point vibrational state and $\alpha_1$ is the direction of the spherical nucleus. For simplicity, we use $\alpha_1=0$. Consequently, a partial fusion cross section is found by averaging over values of the vibrational states $\beta_2$ and $\beta_3$  of the spherical nucleus:
\begin{eqnarray}
\langle \sigma_{\rm fus} (E_{\rm c.m},\ell) \rangle_{\beta^{(1)}_i} &=& 
\int^{\beta_{2+}}_{-\beta_{2+}}d\beta^{(1)}_2\int^{\beta_{3-}}_{-\beta_{3-}}d\beta^{(1)}_3 g(\beta^{(1)}_{2},\beta^{(1)}_{3})
\nonumber\\
&\times& \sigma_{\rm fus}(E_{\rm c.m},\beta^{(1)}_{2},\beta^{(1)}_{3},\ell) 
\label{vibr}
\end{eqnarray}

\begin{table}[ht]
\caption{Deformation parameters $\beta_2$ and $\beta_3$ of the ground states and of the first excited  2$^+$ and $3^-$ states of nuclei used in the calculations in this work.}
\begin{ruledtabular}
\begin{tabular}{ccccc}
Nucleus & $\beta_2$ \cite{MolNix1995}  & $\beta_3$ \cite{MolNix1995} &
$\beta_{2+}$ \cite{Raman2001} & $\beta_{3-}$ \cite{Spear1989} \\
\hline
$^{51}$V & 0.0 & 0.0 & 0.2 &  0.17\\
$^{248}$Cm & 0.235 & 0.0  & 0.297 & 0.065\\
\end{tabular}
\end{ruledtabular}
\label{defpara}
\end{table}

The surface vibrations are regarded as independent harmonic vibrations and the nuclear radius is considered to be distributed as a Gaussian distribution \cite{Esbensen1981},
\begin{align}
g(\beta_2,\beta_3) = \exp
\left[ -\frac{R_0^2\left[\sum_{\lambda}\beta_{\lambda} Y_{\lambda0}^* (\alpha_1)\right]^2}{2 \sigma_{\beta}^2} \right] (2\pi \sigma_{\beta}^2)^{-1/2}.
\end{align}
\begin{align}
\sigma^2_{\beta_2} = \frac{R_0^2}{4\pi}\sum_{\lambda} \beta_{\lambda}^2.
\end{align}

For the simplicity of presentation of the  formulas (\ref{orien}) and  (\ref{vibr}) obtained by averaging over $\alpha_2$ and $\beta^{(1)}_i$ further we use $\sigma_{\rm fus}(E_{\rm c.m},\ell)= 
\langle \sigma_{\rm fus}(E_{\rm c.m},{\beta^{(1)}_i},\alpha_2,\ell) \rangle_{\alpha_2,\beta^{(1)}_i}$. 

\begin{figure}[ht]
	\includegraphics[width=0.47\textwidth]{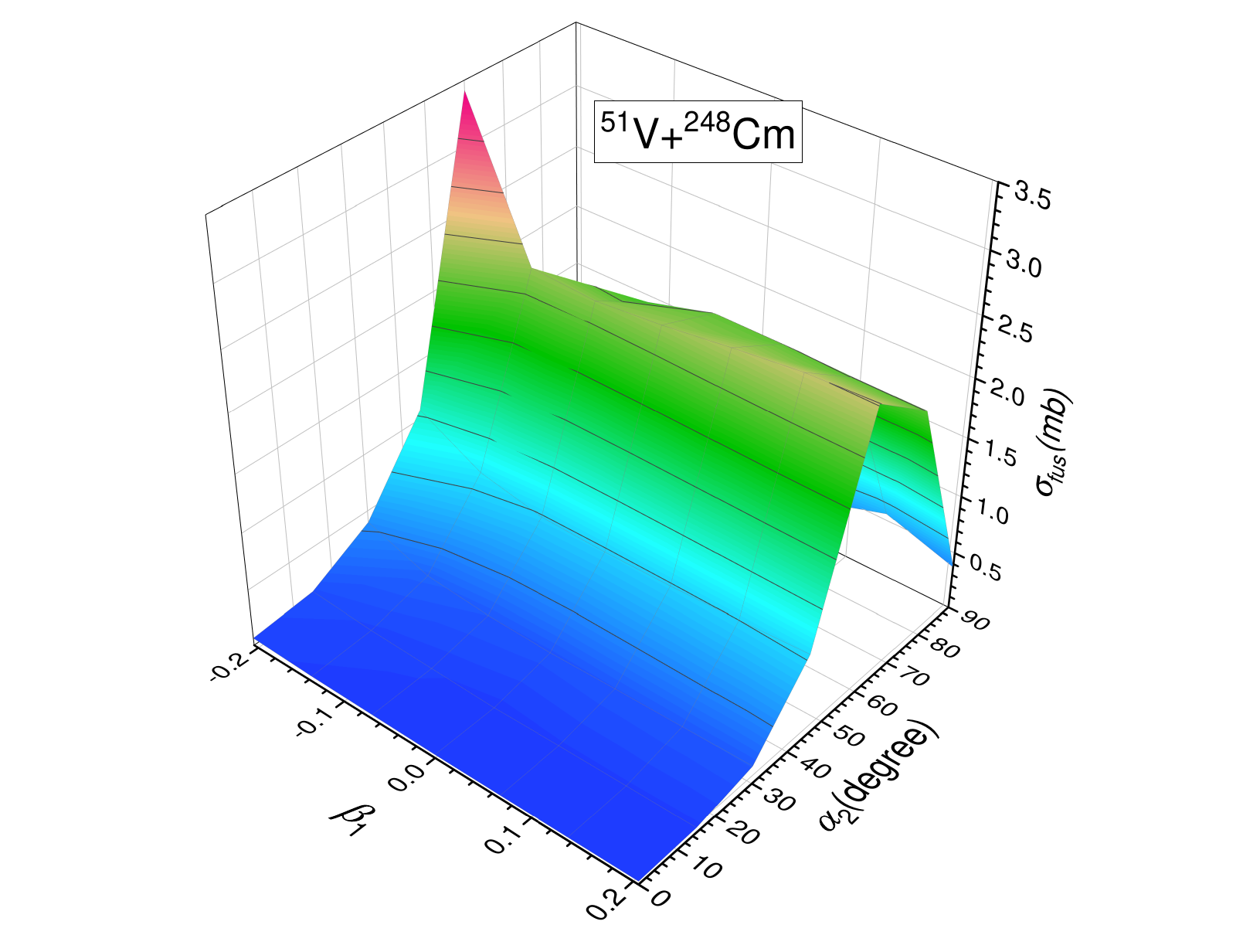}
	\caption{(Color online) The dependence of the cross section of the CN formation in the $^{51}$V+$^{248}$Cm reaction on the orientation angle $\alpha_2$ of the axial symmetry axis of $^{248}$Cm relative to the beam direction and quadrupole deformation values $\beta_{2}$ at surface vibration of the $^{51}$V surface around spherical shape.}
	\label{anglvib}
\end{figure}

The effect of the orientation angle of the axial symmetry axis on the complete fusion is seen from Fig. \ref{anglvib} where the dependence of the cross section of the CN formation in the $^{51}$V+$^{248}$Cm reaction on the orientation angle of the axial symmetry axis of $^{248}$Cm relative to the beam direction and quadrupole deformation values $\beta_{2+}$ at surface vibration is presented. The largest cross section of complete fusion corresponds to the orientation angles 60$^{\circ}-70^{\circ}$ of the target nucleus. 

\begin{figure}[ht]
	\includegraphics[width=0.47\textwidth]{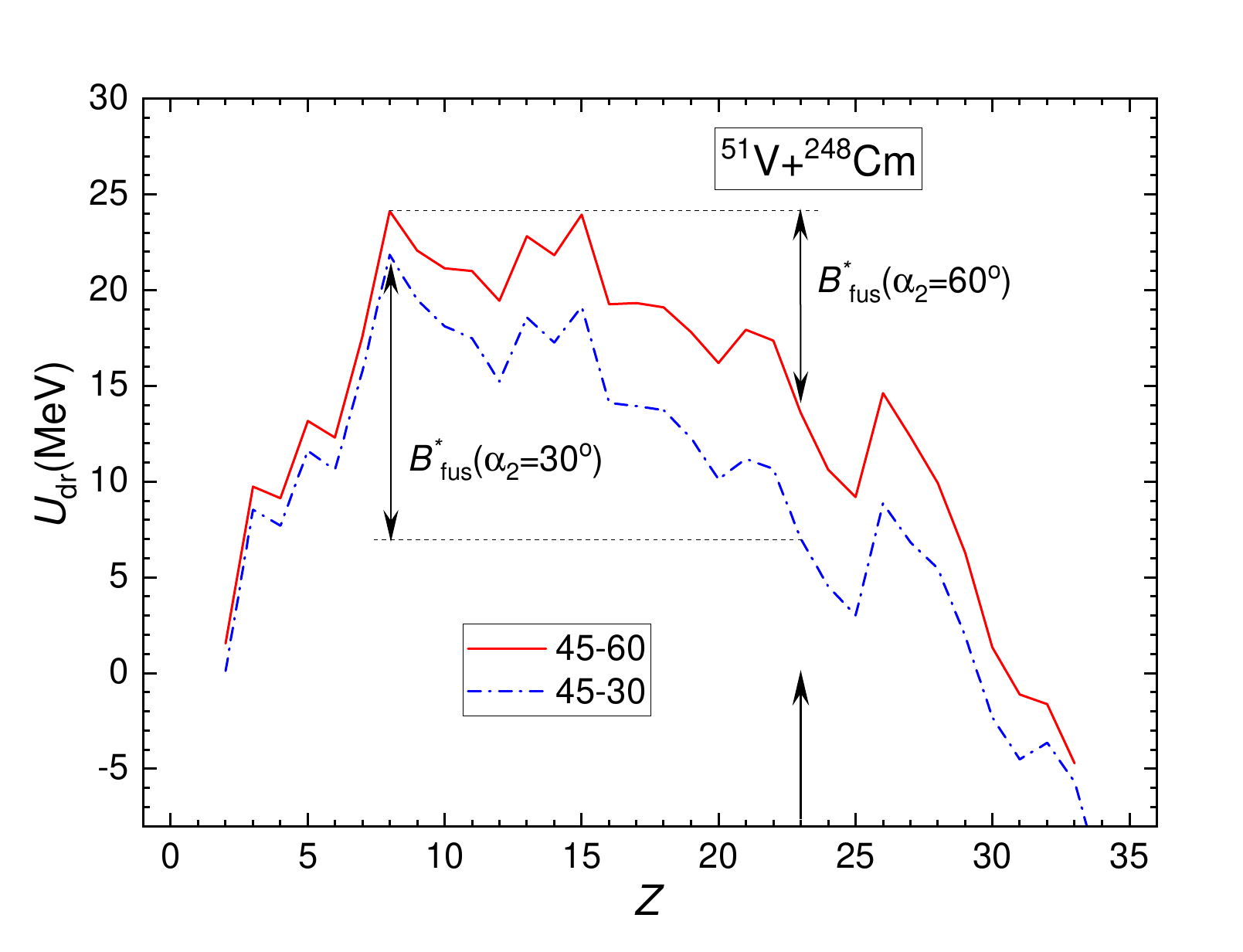}
	\caption{(Color online) The driving potential for the DNS formed in the $^{51}$V+$^{248}$Cm reaction calculated for the orientation angles 30$^{\circ}$ and 60$^{\circ}$ of the axial symmetry axis of $^{248}$Cm relative to the beam direction. The intrinsic fusion barrier $B^*_{\rm fus}$ is shown for the charge asymmetry of the entrance channel $Z=23$.}
	\label{driving}
\end{figure}

The reason of the increasing the cross section of complete fusion at the orientation angles 60$^{\circ}-70^{\circ}$ of the target nucleus  $^{248}$Cm is related with the decrease of the intrinsic fusion barrier $B^*_{\rm fus}$ at these values of $\alpha_2$. The values of the $B^*_{\rm fus}$ determined from the peculiarities of the driving potential are 10.67 and 6.20 MeV for orientation angles $\alpha_2=30^{\circ}$ and 60$^{\circ}$, respectively (See Fig. \ref{driving}.).

\begin{figure}[ht]
	\includegraphics[width=0.47\textwidth]{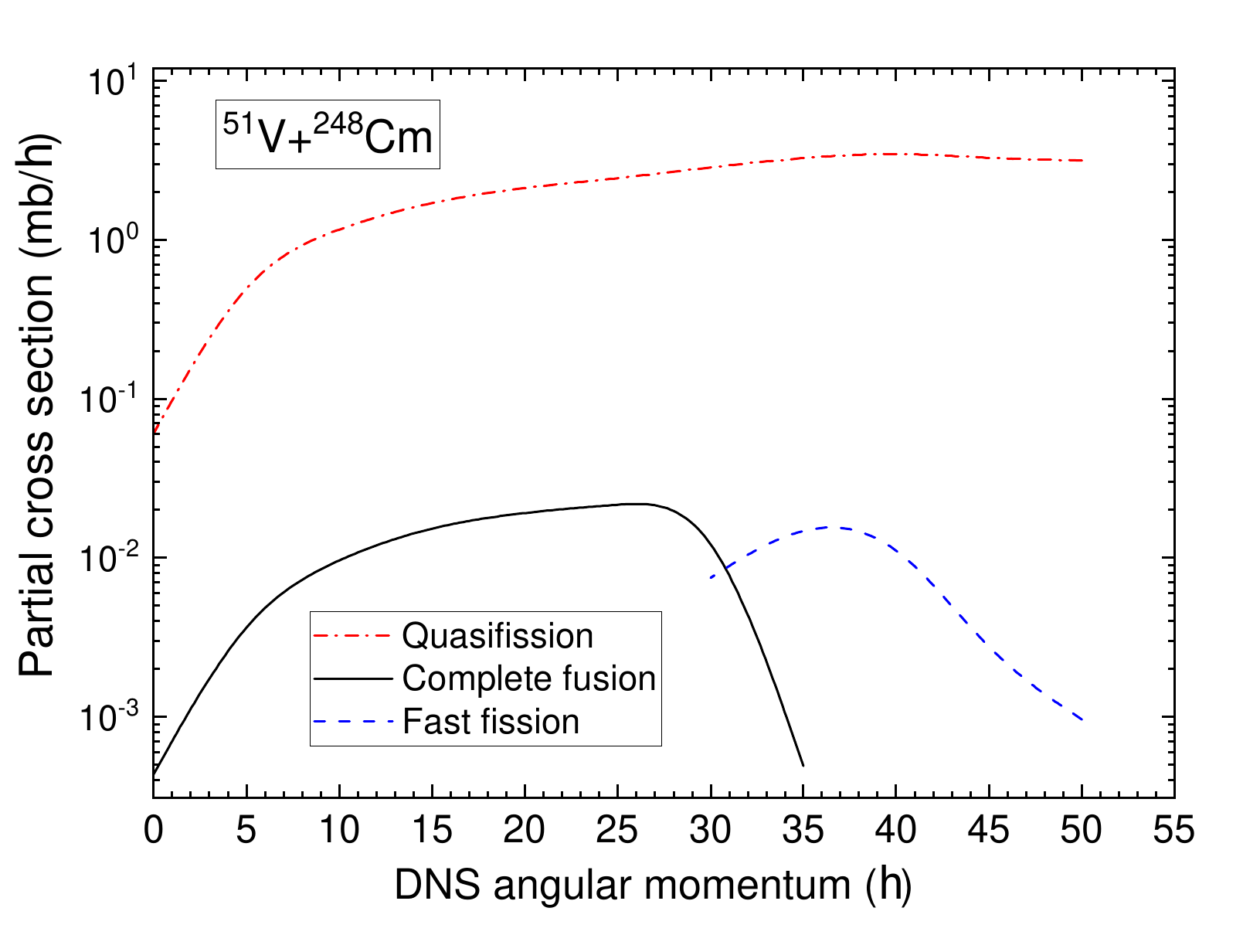}
	\caption{(Color online) Comparison of the partial cross sections of the quasifission (dot-dashed curve), complete fusion (solid curve) and fast fission (dashed curve) calculated for the collision energy $E_{\rm c.m.}=236$ MeV as a function of the DNS angular momentum $\ell$.
}
	\label{partialcross}
\end{figure}

The complete fusion cross section is calculated as a sum of its partial cross sections:
\begin{equation}\label{compfus}
\sigma_{\rm fus}(E_{\rm c.m.})=\sum_{\ell_m}^{\ell_f}\sigma_{\rm cap}(E_{\rm c.m.},\ell)P_{\rm CN}(E_{\rm c.m.},\ell).
\end{equation}
For the $^{51}$V+$^{248}$Cm reaction for all considered collisions we obtained $\ell_m=0$. The initial maximum value of $\ell_d$ has been taken equal to 50 and the calculations of the dynamics of collision gives different values for complete fusion and quasifission processes (see Figs. \ref{partialcross}, \ref{PcnEL}, \ref{PartQfis}, \ref{PartFus}). 

\subsection{Quasifission cross section}
\label{Qfissec}
The quasifission process is an alternative way to the complete fusion in the evolution of the DNS formed at capture of the projectile by the target-nucleus. Therefore,  the quasifission probability is estimated by the expression $P_{\rm qf}(E_{\rm c.m.},\ell)=1-P_{\rm CN}(E_{\rm c.m.},\ell)$. Consequently, the partial quasifission cross section is determined by the product of capture cross section $\sigma_{\rm cap}(E_{\rm c.m.},\ell)$ and  quasifission probability $P_{\rm qf}$:
\begin{eqnarray}\label{qfiss}
\sigma_{\rm qfis}(E_{\rm c.m.},\ell)=\sigma_{\rm cap}(E_{\rm c.m.},\ell) P_{\rm qf}(E_{\rm c.m.},\ell).
\end{eqnarray}
It should be noted that the quasifission process takes place at all values of the orbital angular momentum $\ell$ leading to capture, including head-on collisions of the projectile and target-nuclei.

\subsection{Fast fission cross section}
\label{Fastfissec}
The fast fission process related with the mononucleus formed from the DNS, which has been survived against quasifission, and it being to be transformed to the CN. The CN stability depends on the fission barrier which is calculated as a sum of the parameterized macroscopic fission barrier $B_{fis}^{m}(\ell)$  depending on the angular momentum $J$ \cite{Sierk1986} and  the microscopic (shell) correction $\Delta E_{sh}$ 
\begin{equation}
\label{fissb} B_{fis}(\ell,T)=c \ B_{fis}^{m}(\ell)-h(T) \ q(\ell) \ \Delta E_{sh},
\end{equation}
where $c=1$; $h(T)$ and $q(\ell)$ represent the damping functions of the nuclear shell correction $\Delta E_{sh}$ by the increase of the excitation energy $E^*_{\rm CN}$  and $\ell$ angular momentum, respectively \cite{aglio2012}:
\begin{equation}
h(T) = \{ 1 + \exp [(T-T_{0})/ d]\}^{-1}
\label{hoft}
\end{equation}
and
\begin{equation}
q(\ell) = \{ 1 + \exp [(\ell-\ell_{1/2})/\Delta \ell]\}^{-1},
\label{hofl}
\end{equation}
where, in Eq. (\ref{hoft}), $T=\sqrt{E^*_{\rm CN}/a}$ is the effective nuclear temperature depending on the excitation energy $E^*_{\rm CN}$ and the level density parameter $a=A_{\rm tot}/10$, $d= 0.3$ MeV is the rate of washing out the shell corrections with the temperature, and $T_0=1.16$~MeV is the value at which the damping factor $h(T)$ is reduced by 1/2; analogously, in Eq. (\ref{hofl}), $\Delta \ell=3\hbar$ is the rate of washing out the shell corrections with the angular momentum, and $\ell_{1/2}=20\hbar$ is the value at which the damping factor $q(\ell)$ is reduced by 1/2. 

The fast fission occurs when the orbital angular momentum has a value  $\ell > \ell_f$, where $\ell_f$ is its value leading to the negligible small fission barrier $B_{fis}(\ell,T)$ as a function of $T$. 

Due to including dependence of the fission barrier  $B_{fis}(\ell,T)$ of the CN on its angular momentum $\ell$ and temperature, the cross section of the fast fission process can be calculated by following equation:
\begin{eqnarray}\label{fastfission}
\sigma_{\rm fast.fis}(E_{\rm c.m.})=\sum_{\ell_f}^{\ell_d}\sigma_{\rm cap}(E_{\rm c.m.},\ell)P_{\rm CN}(E_{\rm c.m.},\ell).
\end{eqnarray}
There is two differences in the formation of the quasifission and fast fission products: 1) the quasifission can take place at all values the orbital angular momentum while the fast fission occurs only at its values causing full damping of the fission barrier $B_{fis}$; 2) the quasifission is a breakup of the DNS formed at capture while  the fast fission is the fission of the mononucleus which has survived against quasifission and going to be transformed to the CN.

\subsection{Survival probability}
\label{Surv}
The evaporation residue (ER) cross sections at the given value of the CN excitation energy $E^*_{x}$ is calculated as a sum of the partial cross sections found for the angular momentum $\ell$ of the system:
\begin{eqnarray}\label{erpar}
    \sigma^{(x)}_{ER}(E^*_{x})=\sum^{\ell_d}_{0}\sigma^{(x)}_{ER}(E^*_{x},\ell),
\end{eqnarray}
where, $\sigma^{(x)}_{ER}(E^*_{x},\ell)$ describes the cross section of the particle emission from the intermediate nucleus with the excitation energy $E^*_{x}$ at each step $x$ of the de-excitation cascade by the formula \cite{Nasirov2005, Mandaglio2012}:
\begin{equation}\label{ercs}
    \sigma^{(x)}_{ER}(E^*_{x},\ell)=\sigma^{x-1}_{ER}(E^*_{x-1},\ell)W^{(x)}_{\rm sur}(E^*_{x-1},\ell).
\end{equation}
Here, $\sigma^{(x-1)}_{ER}(E^*_{x-1},\ell)$ is the partial cross section of the  formation of the intermediate excited nucleus at the $(x-1)$th step, and obviously,
\begin{equation}
\sigma^{(0)}_{ER}(E^*_{\rm CN},\ell)=\langle \sigma_{\rm fus} (E_{\rm c.m},\ell) \rangle_{\beta^{(1)}_i},
\end{equation}
which is calculated by (\ref{orien}) and (\ref{vibr}); $W^{(x)}_{\rm sur}(E^*_{x-1},\ell)$ is the survival probability of the $x$th intermediate nucleus against fission along the de-excitation cascade of CN. The survival probability $W^{(x)}_{\rm sur}(E^*_{x-1},\ell)$ is calculated by the statistical model implanted in KEWPIE2 \cite{Kewpie2}, which is dedicated to the study of the evaporation residues at the synthesis of SHE.

The probability of the SHE depends on the competition between the fission and neutron evaporation which is determined by the survival probability $W^{(x)}_{\rm sur}(E^*_{x-1},\ell)$. It has been calculated by the statistical model 
\cite{Kewpie2}, where Weisskopf approximation \cite{Weisskopf1937} is used to 
calculate neutron emission width:
\begin{eqnarray}\label{WidthN}
    \Gamma_n=\frac{(2S_n+1)\mu_n}{\pi^2\hbar^2}\int\limits_0^{E^*_{CN}-B_n}\frac{\sigma^n_{inv}(\epsilon_n)\rho_B(E^*_B)\epsilon_nd\epsilon_n}{\rho_{CN}(E^*_{CN})},
\end{eqnarray}
where $\rho_{CN}(E^*_{CN})$ is the level density of the intermediate nucleus, $E^*_B$ is the excitation energy of the residual nucleus after the emission of a neutron, $B_n$ is the binding energy of the neutron with the residue nucleus, $\mu_n=m_n  (M_{CN}-m_n)/M_{CN}$ is the reduced mass of the system consisting of neutron and evaporation residue nucleus, where $m_n$ and $M_{CN}$ are masses of neutron and CN, respectively. $S_n$ and $\epsilon_n$ denote the intrinsic spin of the neutron and its kinetic energy, $\sigma^n_{inv}$ is the cross section for the time-reversed reaction.

In the KEWPIE2 code, the fission-decay width is estimated within the standard Bohr–Wheeler transition-state model \cite{Wheeler1939}:
\begin{eqnarray}\label{WidthFis}
\Gamma_f^{BW}&=&\frac{1}{2\pi\rho_{CN}^{gs}(E^*_{CN},J_{CN})}\int\limits^{E^*_{CN}-B_f}_{0}\rho_C^{sd}(E^*_{sd},J_{CN})\nonumber\\
&\times& T_{\rm fiss}(\epsilon_f)d\epsilon_f,
\end{eqnarray}
here, the excitation energy at the saddle point is equal $E^*_{sd}=E^*_{CN}-B_f-\epsilon_f$, $\rho_{CN}^{gs}(E^*_{CN}, J_{CN})$ and $\rho_C^{sd}(E^*_{sd}, J_{CN})$ are the level densities of the nucleus at the ground-state and saddle-point deformation, $J_{CN}$ is the total angular momentum of CN.

The KEWPIE2 code \cite{Kewpie2} utilizes the enhanced state-density equation, initially suggested in Ref.\cite{Grossjean1985}, to calculate different decay widths. The penetration factor, $T_{\rm fiss}(\epsilon_f)$, corresponds to the Hill-Wheeler transmission coefficient \cite{Wheeler1953}. The level-density parameter has been taken from the work which was recently proposed by Nerlo-Pomorska et al. \cite{Pomorska2006}. In this work, the estimates for the level-density parameter obtained for different deformations terms are fitted by a liquid-drop type formula, which is expressed in the following form:
\begin{eqnarray}\label{lvldensity}
a=0.092A&+&0.036A^{2/3}\mathfrak{B}_s+0.275A^{1/3}\mathfrak{B}_k\nonumber\\
&&-0.00146\frac{Z^2}{A^{1/3}}\mathfrak{B}_c,     
\end{eqnarray}
where $\mathfrak{B}_s$ is the surface term, $\mathfrak{B}_k$ is the curvature term and $\mathfrak{B}_c$ is the Coulomb term for a deformed nucleus \cite{Hasse}, and they are given in the following forms: 
\begin{eqnarray}
    \mathfrak{B}_s=1+\frac{2}{5}\alpha_2^2-\frac{4}{105}\alpha_2^3-\frac{66}{175}\alpha_2^4, \nonumber\\
    \mathfrak{B}_k=1+\frac{2}{5}\alpha_2^2+\frac{16}{105}\alpha_2^3-\frac{82}{175}\alpha_2^4,\\
    \mathfrak{B}_c=1-\frac{1}{5}\alpha_2^2-\frac{4}{105}\alpha_2^3+\frac{51}{245}\alpha_2^4, \nonumber
\end{eqnarray}
here $\alpha_2=\sqrt{5/4\pi}\beta_2$, where $\beta_2$ refers to the quadrupole deformation parameter. 

The shell-correction effects decrease as the excitation energy rises. To incorporate this damping effect, Ignatyuk's prescription \cite{Ignatyuk75} was used in the calculation, which considers the level density parameter to be dependent on the excitation energy. In the ground state, one has the following explicit expression:
\begin{eqnarray}\label{Ignlevden}
a_{gs}(E^*_x)=a\left[1+\left(1-e^{-E^*_x/E_d}\right)\frac{\Delta E_{sh}}{E^*_x}\right],
\end{eqnarray}
where the default value of the shell-damping energy $E_d$ has been fixed at 19.0 MeV on the base of our previous calculations of the decay for the different reactions. The fission barrier consists of two parts calculated by the macroscopic liquid-drop model and microscopic shell correction energy \cite{Giardina2018}. The liquid-drop fission barrier is estimated by using the Lublin-Strasbourg Drop model \cite{Lubin2009}. The ground-state shell correction energies and the parameterizations for the liquid-drop fission barrier are using the mass table of M\"{o}ller et al. \cite{MolNix1995}.

\section{Results and discussions}
\label{discuss}
\begin{figure}[ht]
	\includegraphics[width=0.48\textwidth]{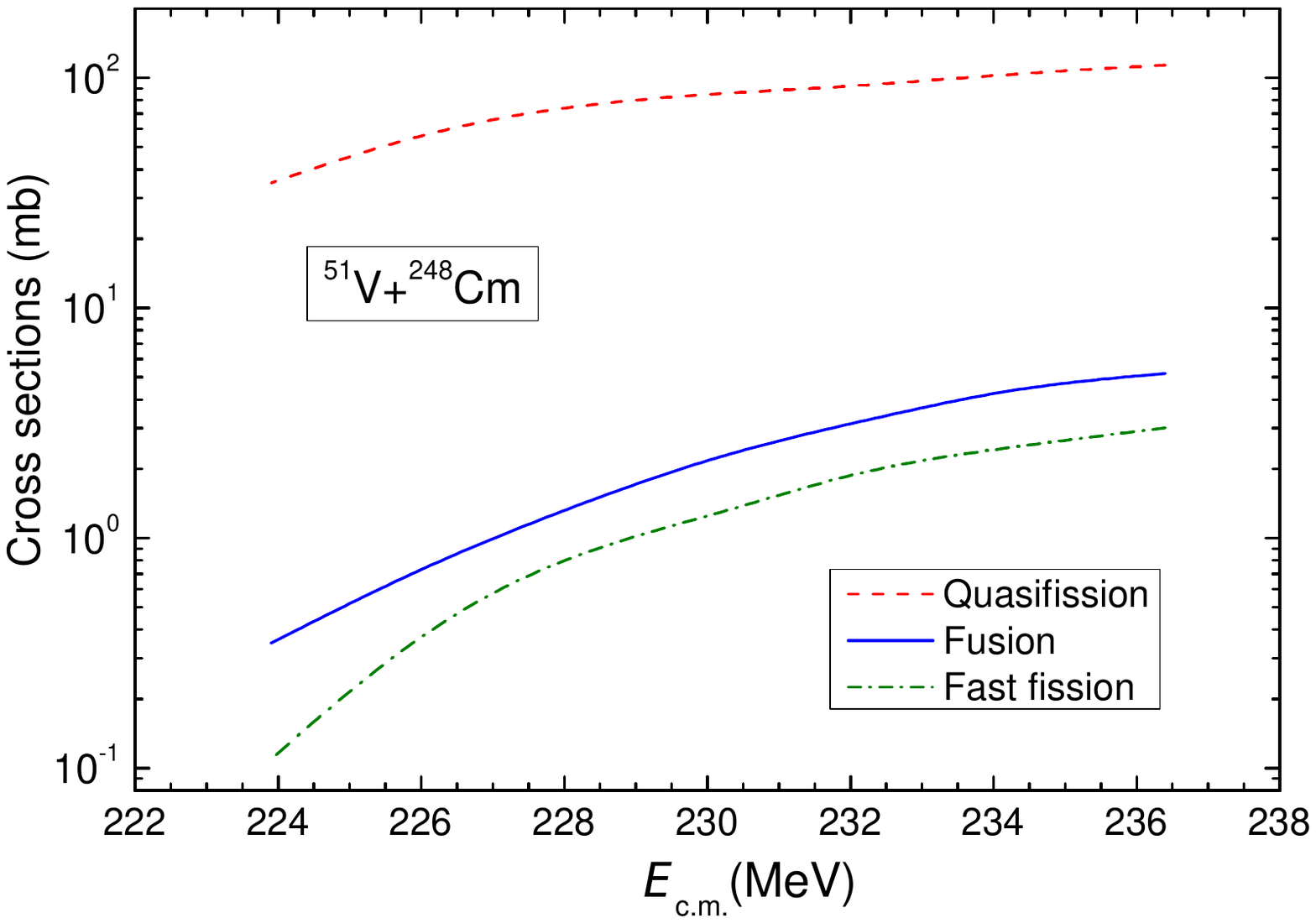}
	\caption{(Color online) Theoretical cross sections of the quasifission (red dashed curve), complete fusion (solid blue curve), and fast fission (dot-dashed green curve) for the $^{51}$V+$^{248}$Cm reaction as a function of $E_{\rm c.m.}$.}
	\label{TotalCross}
\end{figure}

The results of the total cross sections of the quasifission, complete fusion, and fast fission processes obtained in this work are presented in Fig. \ref{TotalCross} by the red dashed, blue solid, and green dot-dashed curves, respectively. It is seen that quasifission is the dominant channel since the intrinsic barrier $B^*_{\rm fus}$ causing a hindrance to complete fusion is sufficiently large and quasifission barrier  $B_{\rm qf}$ providing the stability of the DNS is small. Therefore, the fusion probability is small too.

\begin{figure}[ht]
	\centering
	\resizebox{0.48\textwidth}{!}
	{\includegraphics[width=7.8cm,height=6cm]{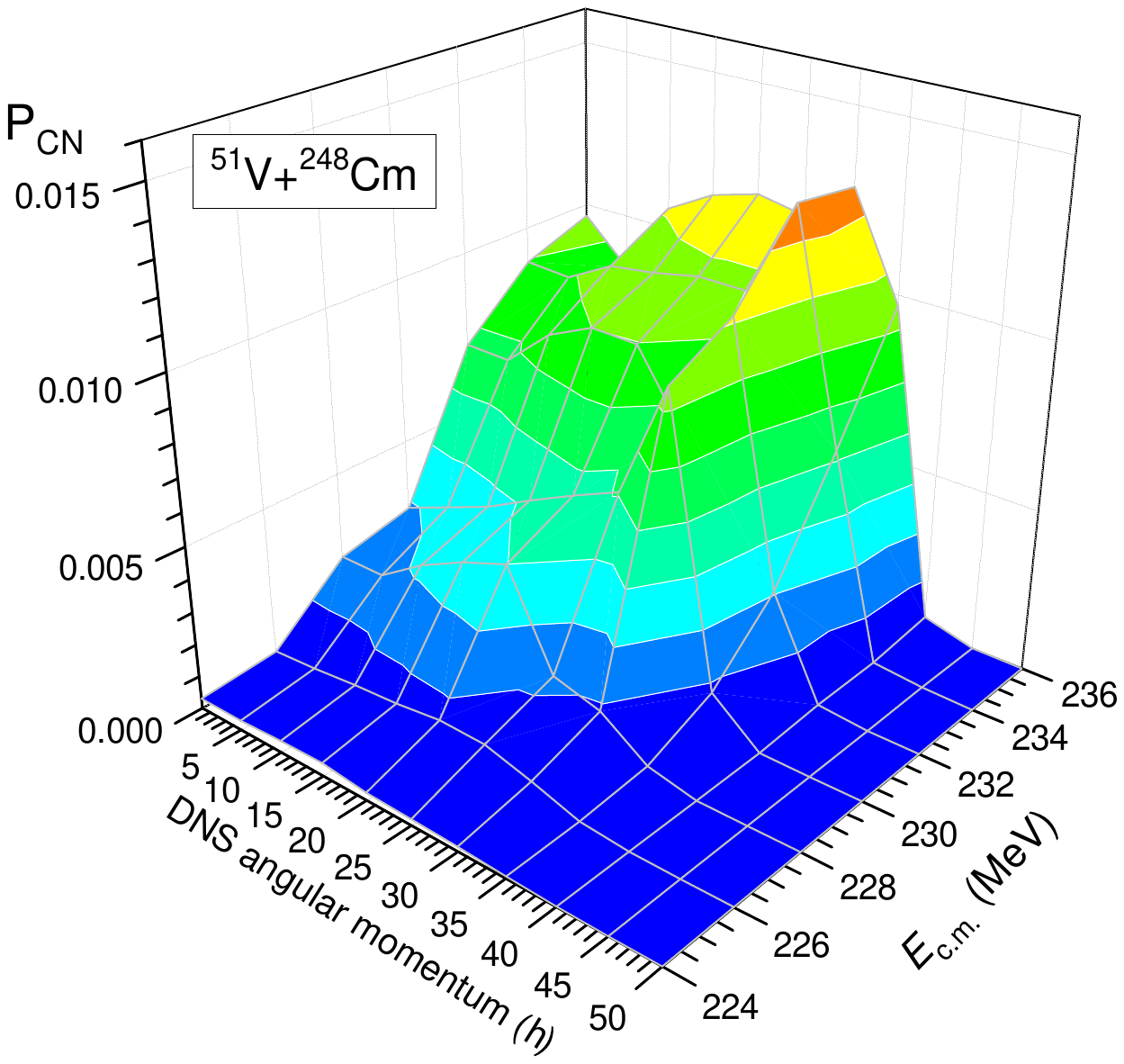}}
	\caption{(Color online) Dependence of the fusion probability $P_{\rm CN}$ calculated for the $^{51}$V+$^{248}$Cm reaction as a function $E_{\rm c.m.}$ and angular momentum.}
	\label{PcnEL}
\end{figure}
\begin{figure}[ht]
	\centering
	\resizebox{0.48\textwidth}{!}
	{\includegraphics[width=7.8cm,height=6cm]{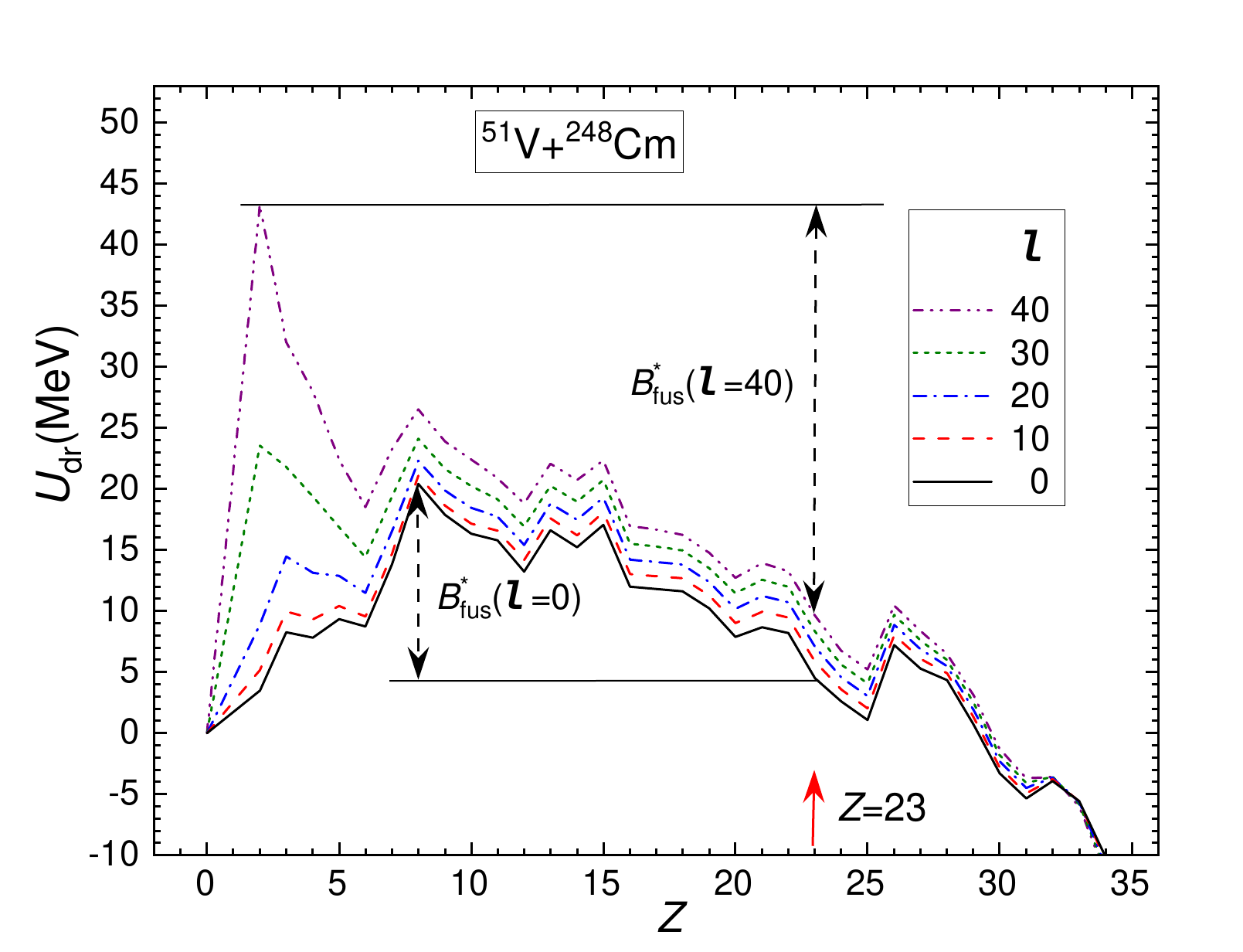}}
	\caption{(Color online) The dependence of the driving potential calculated for the $^{51}$V+$^{248}$Cm reaction as a function of the orbital angular momentum $\ell$. The intrinsic fusion barrier $B^*_{\rm fus}$ is shown for the charge asymmetry of the entrance channel $Z=23$.
  The results are presented for the orientation angle of $^{248}$Cm $\alpha_2=60^{\circ}$ and $\beta_2(1)$=-0.12.}
	\label{BfusL}
\end{figure}

\begin{figure}[ht]
	\centering
	\resizebox{0.48\textwidth}{!}
	{\includegraphics[width=7.8cm,height=6cm]{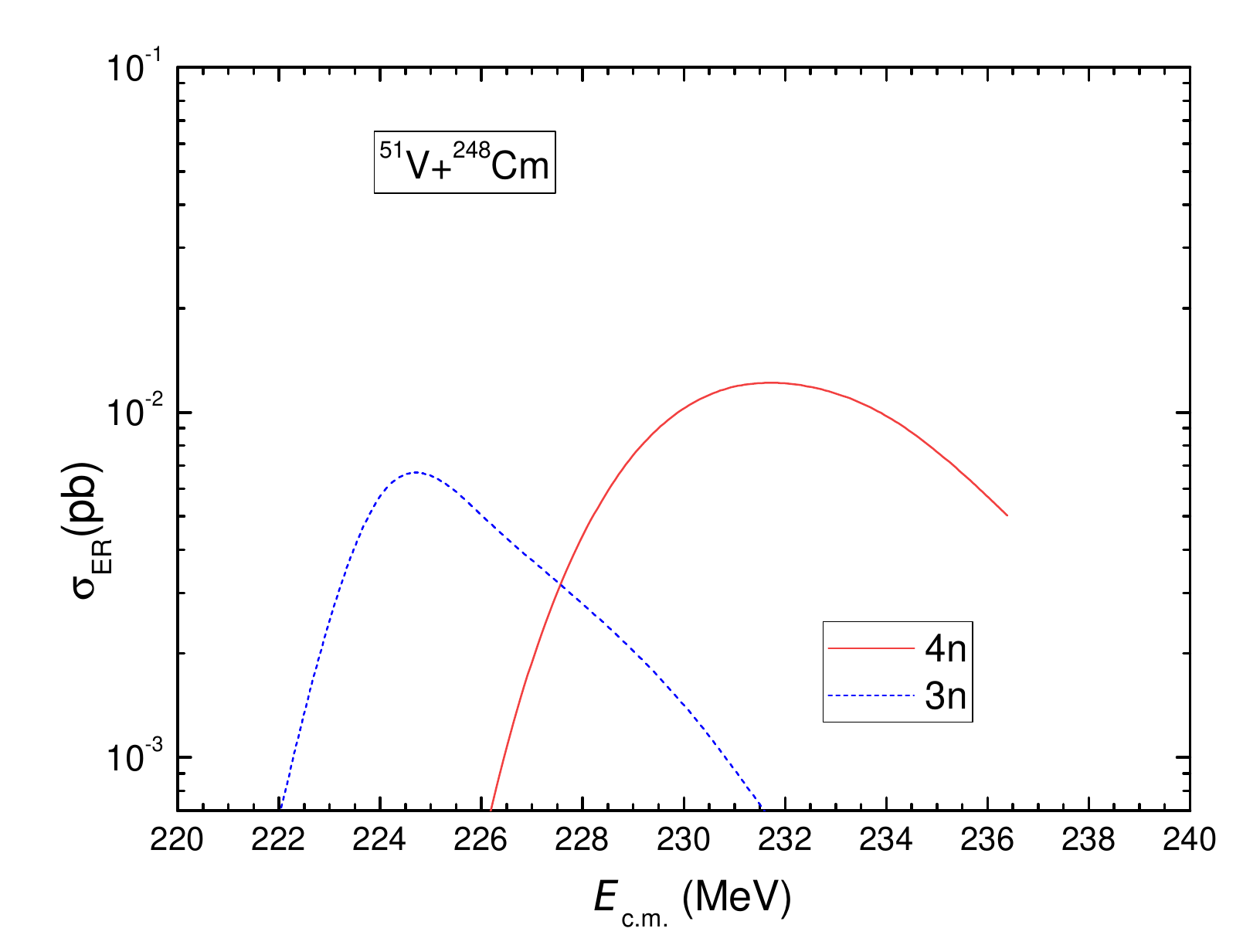}}
	\caption{(Color online) Theoretical cross sections of the 3n (dashed blue curve) and 4n (solid red curve) channels of the ER formation  in the $^{51}$V+$^{248}$Cm reaction as a function of $E_{\rm c.m.}$.}
	\label{ERcross}
\end{figure}
\begin{figure}[ht]
	\centering
	\resizebox{0.48\textwidth}{!}
	{\includegraphics[width=7.8cm,height=6cm]{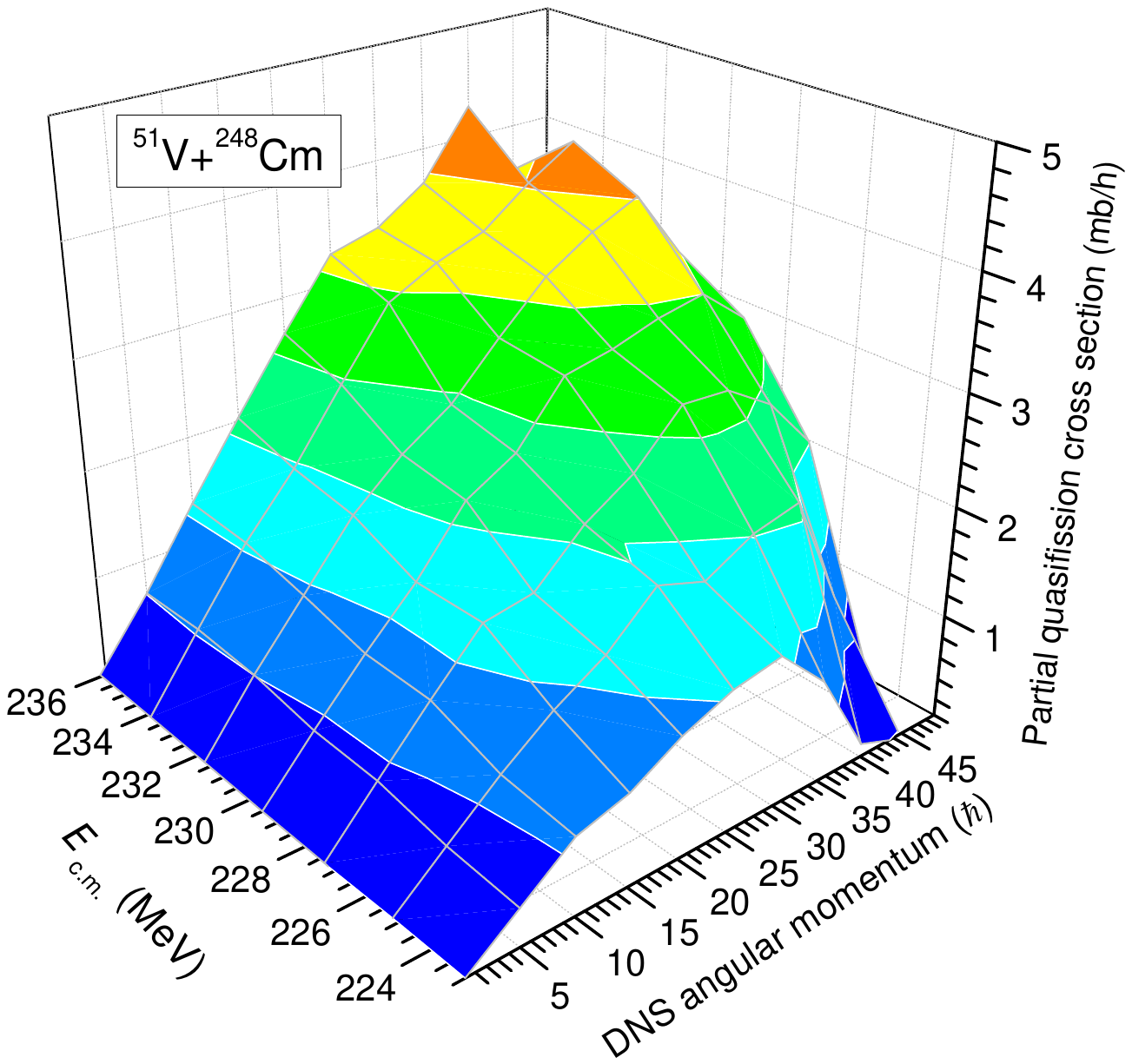}}
	\caption{(Color online) The partial quasifission cross sections calculated for the $^{51}$V+$^{248}$Cm reaction as  a function of $E_{\rm c.m.}$ and orbital angular momentum.}
	\label{PartQfis}   
\end{figure}
The dependence of the fusion probability on the orbital angular momentum and $E_{\rm c.m.}$ is seen in Fig. \ref{PcnEL}. It is calculated as a ratio of the partial cross section of the CN formation $\sigma_{\rm fus}^{(\ell)}$ to the partial capture cross section. It can be seen from Fig. \ref{PcnEL} that the fusion probability is greater by the increase of the beam energy since the DNS excitation energy $E^*_Z$ increases at the fixed intrinsic fusion barrier $B^*_{\rm fus}$. The decrease of the $P_{\rm CN}$ by the increase of angular momentum $L$ is explained by the increase of the $B^*_{\rm fus}$ and decrease of the quasifission barrier $B_{\rm qf}$ as a function of the angular momentum. The maximum value of $P_{\rm CN}$ is obtained in the ranges of the $232 < E_{\rm c.m.} < 236 $ MeV and $ 15 < L < 30 \hbar $. The increase of $B^*_{\rm fus}$ by the increase of  $L$ is related with the increase of the rotational energy of the DNS   (see Fig. \ref{BfusL}). $B^*_{\rm fus}$ is determined as a difference between the maximum value of the driving potential and its value corresponding to the charge and mass number of the colliding nuclei at the given value $L$ (for the details see Refs. \cite{Fazio2005, Nasirov2005}).  

\begin{table*}
\caption{Comparison of the theoretical predictions of the ER cross sections obtained in this work with results presented in Refs. \cite{Zhu2014,Lv2021,Wilczynska2019,Ghahramany2016,Adamian2018,Adamian2020}.}
\label{tabERCS}
\begin{ruledtabular}

\begin{tabular}{cccccc}
Reaction &  $E_{\rm c.m.}$ (MeV) & $3n$ (fb) &  $E_{\rm c.m.}$ (MeV)& $4n$ (fb) & Ref. \\ \hline
&  225 & 6.6 &  232 & 12.3 & (this work) \\
&  227 & 5.5 &  237 & 10.1 & \cite{Zhu2014}  \\
$^{51}$V$+^{248}$Cm & 228 & 2.9 &  236 & 5.9 & \cite{Wilczynska2019}  \\
&    230 & 9.2 & 248 & 1.2 &\cite{Lv2021}  \\
&  232 & 19.5  &  245 & 99.6 & \cite{Ghahramany2016} \\
&  - & -  &  237 & 11.8 & \cite{Adamian2018} \\
\hline
&  -  & -  &  230 & 40 & \cite{Adamian2018}  \\
&  -  & -  &  230 & 18.7 & \cite{Adamian2020}  \\
$^{50}$Ti$+^{249}$Bk& 224   & 11.9  &  236 & 64 & \cite{Zhu2014}  \\
& 226 & 48.2 & 244 & 5.67 &\cite{Lv2021}  \\
& 225 & 29.8 &  233 & 35.6 & \cite{Wilczynska2019}  \\
\end{tabular}
\end{ruledtabular}
\end{table*}
The excitation energy of the formed CN is $35 < E^*_{\rm CN} < 39 $ MeV. The results of the calculation of the ER cross section by the statistical method (KEWPIE2 code \cite{Kewpie2}) presented in Section \ref{Surv} are shown in Fig. \ref{ERcross}. The maximum value of the excitation function for the 4n channel is 12.3 fb (1 fb=$10^{-15}$ b) and it corresponds to the $E_{\rm c.m.} =$ 232 MeV.  The observation of the 3n channel is less probable since it is 6.6 fb which is reached at $E_{\rm c.m.} =$ 225 MeV. In Table \ref{tabERCS}, the results obtained in this work for the ER cross section in the 3n and 4n channels are compared with the corresponding predictions presented in Refs. \cite{Zhu2014, Lv2021, Wilczynska2019, Ghahramany2016, Adamian2018, Adamian2020}. It is seen that our results are close to the predicted values of $\sigma_{\rm ER}$ in Refs. \cite{Zhu2014,Adamian2018}. The values of $E_{\rm c.m.}$ presented in Table \ref{tabERCS} are the energies corresponding to the maximum value of the cross sections of the 3n and 4n channels, which are calculated by using different methods of calculation in the cited papers.  
\begin{figure}[ht]
	\centering
	\resizebox{0.48\textwidth}{!}
	{\includegraphics[width=7.8cm,height=6cm]{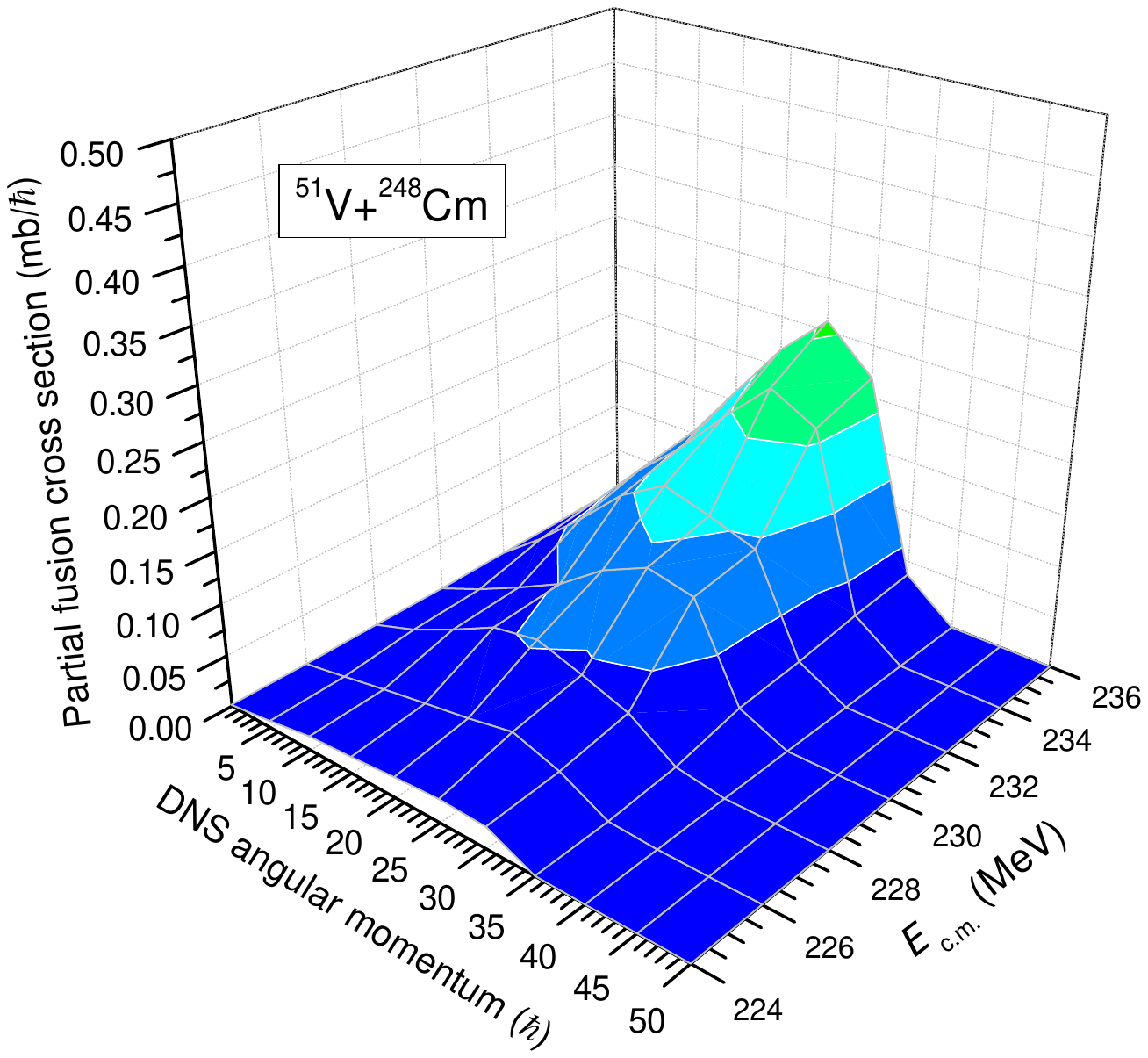}}
	\caption{(Color online) The partial fusion cross sections calculated for the $^{51}$V+$^{248}$Cm reaction as  a function of $E_{\rm c.m.}$ and orbital angular momentum.}
	\label{PartFus}
\end{figure}

One of the reasons the smallest value of the ER cross section is related to the small fusion cross section. It is much smaller than the cross section of quasifission which is the dominant process in the wide range of angular momentum (see Fig. \ref{PartQfis}). The partial fusion cross sections presented in Fig. \ref{PartFus} show that the contribution of the collisions with the orbital angular momentum $L=15-30 \hbar$ at the collision energies $E_{\rm c.m.}=232-236 $ is large. It means that the beam energy being used in the RIKEN experiments is optimal for the synthesis of the new superheavy element with charge number $Z=119$ \cite{Tanaka2022}. The other reason for decreasing the number of synthesis events is the fast fission phenomenon: the mono-nucleus survived against quasifission splits instead of reaching the CN stage. This occurs because of the absence of the fission barrier $B_{\rm fis}$ which is sensitive against its rotational energy and excitation energy. It is well known that the stability of superheavy elements is determined exclusively by the shell effects of their nucleon structure.

\begin{figure}[ht]
	\resizebox{0.49\textwidth}{!}
	{\includegraphics[width=7.8cm,height=6 cm]{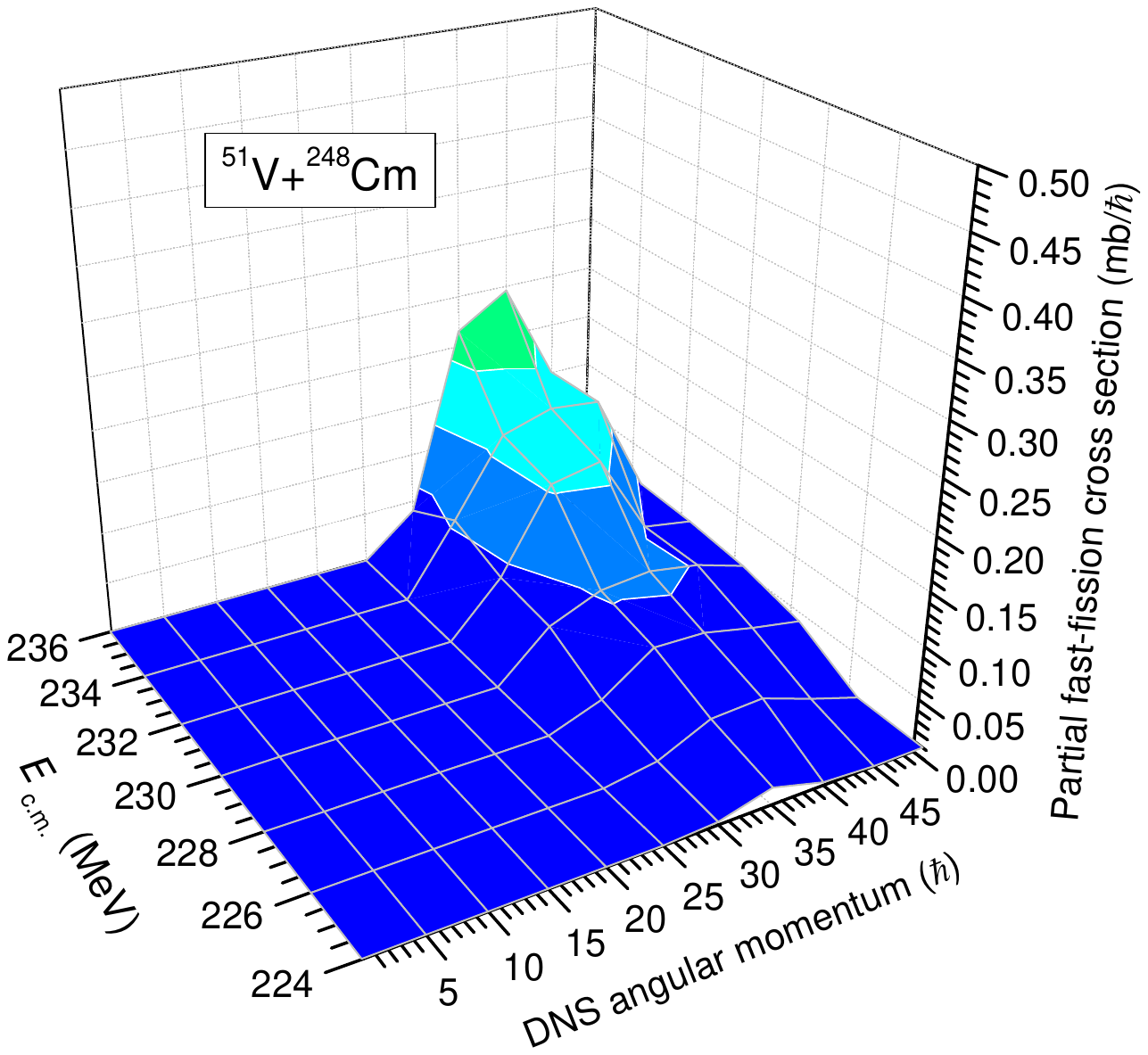}}
	\caption{(Color online) The partial fast fission cross sections calculated for the $^{51}$V+$^{248}$Cm reaction as  a function of $E_{\rm c.m.}$ and orbital angular momentum.}
	\label{PartFasfis}
\end{figure}

The dependence of the fast fission cross section on the orbital angular momentum and collision energy can be seen from Fig. \ref{PartFasfis}. Its maximum values correspond to the collision energies as in the case of the complete fusion but fast fission is strong at the large values of the angular momentum $L=35-45 \hbar$.

\section{Conclusions.}

The ER cross section of 3n and 4n channels related to the synthesis of SHE of the charge number $Z=119$ in the $^{51}$V+$^{248}$Cm reaction has been calculated by the DNS model as a sum of the partial cross sections of the corresponding channels. The angular momentum distribution of the CN is estimated by the dynamical trajectory calculations of the capture probability which is considered as the DNS formation probability. The fusion probability decreases by the increase of the DNS angular momentum due to its influence on the intrinsic fusion barrier $B_{\rm fus}^*$. The range $\alpha_2=60^{\circ}-70^{\circ}$ of the orientation angle of the axial symmetry axis of the deformed target nucleus $^{248}$Cm is favorable for the formation of the CN. The fusion probability decreases at around $\alpha_2=90^{\circ}$ since the number of the partial waves contributing to the capture decreases. Therefore, it is important to calculate the capture cross section dynamically. The 4n channel cross section of the SHE synthesis is larger than the 3n channel cross section and the maximum value of the ER cross section is 12.3 fb at $E_{\rm c.m.}$=232 MeV.
\bibliography{sample}
\end{document}